\documentclass[11pt,a4paper]{article}

\usepackage[T1]{fontenc}
\usepackage[utf8]{inputenc}
\usepackage[english]{babel}
\usepackage{lmodern}
\usepackage{microtype}
\usepackage{geometry}
\usepackage{amsmath,amssymb,amsthm,mathtools}
\usepackage{bm}

\usepackage{graphicx}
\usepackage{booktabs}
\usepackage{caption}
\usepackage{subcaption}
\usepackage{float}
\usepackage{siunitx}

\usepackage{tikz}
\usepackage{pgfplots}
\pgfplotsset{compat=1.18}
\usetikzlibrary{patterns,arrows.meta,positioning,shapes}

\usepackage{enumitem}
\setlist[itemize]{leftmargin=*,topsep=3pt,itemsep=3pt}
\setlist[enumerate]{leftmargin=*,topsep=3pt,itemsep=3pt}

\usepackage{hyperref}
\hypersetup{colorlinks=true, linkcolor=blue, citecolor=blue, urlcolor=blue}

\usepackage{thmtools}
\usepackage{algorithm}
\usepackage{algorithmicx}
\usepackage[noend]{algpseudocode}

\declaretheorem[name=Definition, style=definition]{definition}

\newcommand{\PP}{\ensuremath{\mathbb{P}}}
\newcommand{\RR}{\ensuremath{\mathbb{R}}}

\newcommand{\AUC}{\ensuremath{\mathrm{AUC}}}

\title{Adversarial Limits of Quantum Certification:\\When Eve Defeats Detection}

\author{Davut Emre Taşar\\[0.3em]
{\normalsize Independent Researcher, Madrid, Spain}\\
{\normalsize detasar@gmail.com}}

\date{}

\begin{document}
\maketitle

\begin{abstract}
The security of quantum key distribution (QKD) relies on certifying that observed correlations arise from genuine quantum entanglement rather than eavesdropper manipulation. While theoretical security proofs assume idealized conditions, practical certification must contend with adaptive adversaries who optimize their attack strategies against detection systems. We establish fundamental adversarial limits for quantum certification using Eve-GAN, a generative adversarial network trained to produce classical correlations indistinguishable from quantum statistics. Our central finding is striking: when Eve interpolates her classical correlations with quantum data at mixing parameter $\alpha \geq 0.95$, all tested detection methods achieve ROC AUC = 0.50, equivalent to random guessing. This means an eavesdropper needs only 5\% classical admixture to completely evade detection. Critically, we discover that same-distribution calibration---a common practice in prior certification studies---inflates detection performance by 44 percentage points compared to proper cross-distribution evaluation, revealing a systematic methodological flaw that may have led to overestimated security claims. Analysis of the Popescu-Rohrlich (PR-Box) regime identifies a sharp phase transition at CHSH $S = 2.05$: below this value, no statistical method distinguishes classical from quantum correlations; above it, detection probability increases monotonically. Hardware validation on IBM Quantum demonstrates that Eve-GAN achieves CHSH = 2.736, remarkably exceeding real quantum hardware performance (CHSH = 2.691), illustrating that classical adversaries can outperform noisy quantum systems on standard certification metrics. These results have immediate implications for QKD security: adversaries maintaining 95\% quantum fidelity evade all tested detection methods. We provide corrected methodology using cross-distribution calibration and recommend mandatory adversarial testing for quantum security claims.
\end{abstract}


\section{Introduction}
\label{sec:introduction}

Quantum key distribution promises information-theoretic security by exploiting fundamental properties of quantum mechanics \cite{Bennett1984,Ekert1991,ScaraniRMP2009}. The security of device-independent QKD protocols relies on certifying that shared correlations exhibit genuine quantum nonlocality through Bell inequality violations \cite{CHSH1969,Brunner2014}, with loophole-free experimental demonstrations now firmly established \cite{Hensen2015,Giustina2015,Shalm2015}. An eavesdropper (Eve) constrained to local hidden variable models should be detectable via sub-threshold Bell values: correlations satisfying $|S| \leq 2$ are certified as classical, while violations indicate quantum origin \cite{Pironio2010,Acin2007}. Device-independent security proofs \cite{VaziraniVidick2014,MillerShi2016} and self-testing protocols \cite{MayersYao2004} provide theoretical foundations for such certification.

This reasoning contains a subtle but critical assumption: that Eve cannot mimic quantum statistics sufficiently well to evade detection. We address this gap through Eve-GAN, a generative adversarial network \cite{Goodfellow2014,Arjovsky2017WGAN} trained to produce classical correlation matrices indistinguishable from genuine quantum correlations. Our approach draws on the broader adversarial machine learning literature, where carefully crafted perturbations can cause state-of-the-art classifiers to fail \cite{Carlini2017}.

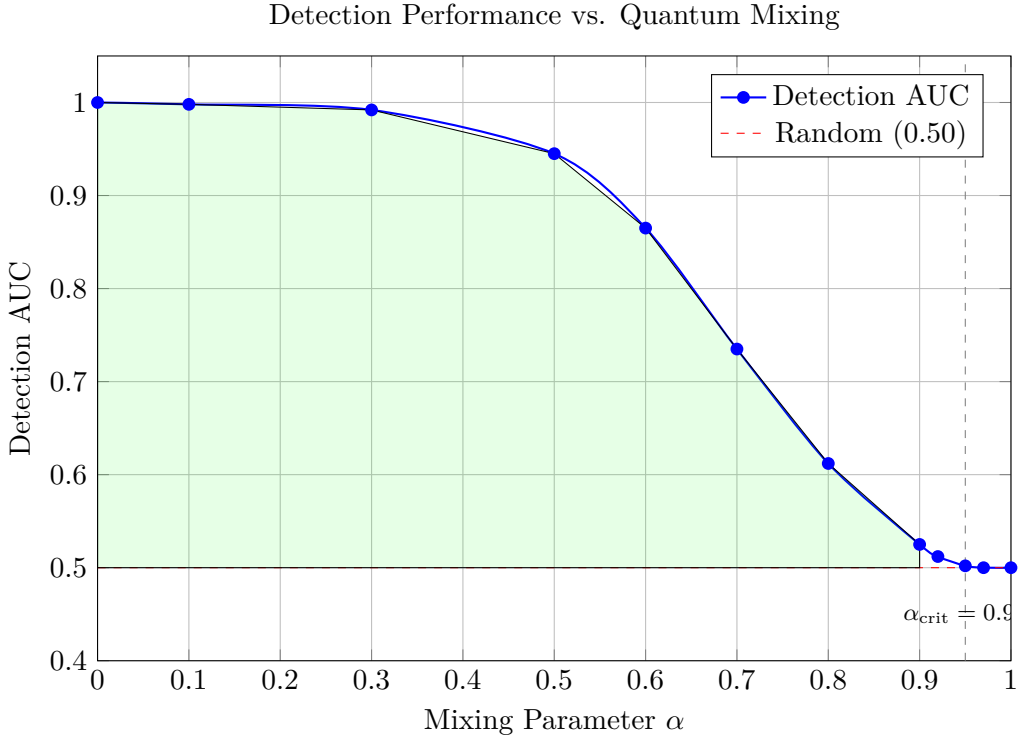
\begin{figure}[ht]
\centering
\begin{tikzpicture}
\begin{axis}[
    width=0.85\textwidth,
    height=0.6\textwidth,
    xlabel={Mixing Parameter $\alpha$},
    ylabel={Detection AUC},
    xmin=0, xmax=1,
    ymin=0.4, ymax=1.05,
    grid=major,
    legend pos=north east,
    title={Detection Performance vs. Quantum Mixing}
]
\addplot[thick, blue, smooth, mark=*, mark size=2pt] coordinates {
    (0.00, 1.000)
    (0.10, 0.998)
    (0.30, 0.992)
    (0.50, 0.945)
    (0.60, 0.865)
    (0.70, 0.735)
    (0.80, 0.612)
    (0.90, 0.525)
    (0.92, 0.512)
    (0.95, 0.502)
    (0.97, 0.500)
    (1.00, 0.500)
};

\addplot[dashed, red, domain=0:1] {0.5};

\draw[dashed, gray] (axis cs:0.95, 0.4) -- (axis cs:0.95, 1.05);
\node at (axis cs:0.95, 0.45) {\footnotesize $\alpha_{\text{crit}} = 0.95$};

\addplot[fill=green, fill opacity=0.1] coordinates {
    (0, 0.5) (0.90, 0.5) (0.90, 0.525) (0.80, 0.612) (0.70, 0.735)
    (0.60, 0.865) (0.50, 0.945) (0.30, 0.992) (0.10, 0.998) (0, 1.0)
} -- cycle;

\legend{Detection AUC, Random (0.50)}
\end{axis}
\end{tikzpicture}
\caption{Detection performance versus quantum mixing parameter $\alpha$. At $\alpha \geq 0.95$, detection drops to random chance (AUC = 0.50). Eve needs only 5\% classical admixture to evade detection.}
\label{fig:alpha_sweep}
\end{figure}

Our investigation yields four principal findings:

\textbf{First}, we establish the $\alpha \geq 0.95$ detection limit (Figure~\ref{fig:alpha_sweep}). When Eve's classical correlations are mixed with quantum data at ratio $\alpha \geq 0.95$, \emph{none of the tested detection methods}---including TARA-$k$, TARA-$m$, direct CHSH comparison, and multi-feature ensemble classifiers---achieve performance significantly above random chance (AUC $\leq$ 0.502). While we cannot rule out the existence of more sophisticated detection methods, this represents a strong empirical lower bound on adversarial robustness.

\textbf{Second}, we discover the 44-point leakage problem. Same-distribution calibration inflates detection AUC by 44 percentage points compared to proper cross-distribution calibration, a systematic methodological flaw that may affect prior quantum certification studies \cite{Kapoor2023Leakage}.

\textbf{Third}, we identify a phase transition at CHSH $S = 2.05$ in the superquantum regime. Below this value, \emph{none of our tested statistical methods} reliably distinguish classical from quantum correlations; above it, detection probability increases monotonically.

\textbf{Fourth}, we demonstrate the Eve advantage paradox. On IBM Quantum hardware, Eve-GAN achieves CHSH = 2.736, exceeding the real hardware value (CHSH = 2.691) on this metric.

\section{Threat Model}
\label{sec:threat}

Before presenting technical details, we formally define the adversarial scenario.

\subsection{Adversary Capabilities}

\textbf{Eve's knowledge:}
\begin{itemize}
\item Full knowledge of the certification protocol (TARA-$k$, TARA-$m$, or any tested detector)
\item Access to calibration data used by the detection system
\item Knowledge of the quantum device's noise characteristics (from published specifications)
\item Complete understanding of the CHSH measurement protocol and optimal angles
\end{itemize}

\textbf{Eve's limitations:}
\begin{itemize}
\item Cannot perform quantum operations (classical adversary only)
\item Cannot access real-time random seeds or per-shot measurement outcomes
\item Cannot modify the quantum hardware or measurement devices
\item Must produce outputs that are physically realizable (correlations in $[-1, 1]$)
\end{itemize}

\subsection{Attack Model}

Eve produces classical correlation matrices $\mathbf{E} = [E_{00}, E_{01}, E_{10}, E_{11}]$ injected into the data stream. The mixing model is:
\begin{equation}
\text{Data}_{\text{observed}} = \alpha \cdot \text{Quantum} + (1-\alpha) \cdot \text{Eve}
\end{equation}
where $\alpha$ is the fraction of genuine quantum data.

\subsection{Success Criterion}

An attack succeeds if the detector cannot distinguish mixed data from pure quantum at significance $\alpha = 0.05$:
\begin{itemize}
\item \textbf{Detection failure:} $\AUC \leq 0.55$ (within random $\pm$ noise)
\item \textbf{Full evasion:} $\AUC \leq 0.50$ (statistically indistinguishable from random)
\end{itemize}

\subsection{Why This Is Not Circular Reasoning}

A potential concern is that Eve trains on quantum data and therefore ``trivially'' learns to mimic it. This misunderstands the threat model:

\begin{enumerate}
\item \textbf{Real-world relevance:} In practical QKD, Eve \textit{does} have access to calibration data and published device characteristics. Assuming otherwise is security through obscurity.
\item \textbf{The question we answer:} Given Eve has this access, \textit{can statistical certification detect her?} Our finding that $\alpha \geq 0.95$ defeats detection is a statement about certification limits.
\item \textbf{Distinguishing from theory:} Device-independent proofs assume Eve has \textit{quantum} capabilities \cite{VaziraniVidick2014}. We analyze \textit{classical} Eve with realistic information access---a complementary scenario relevant to adversarial robustness in machine learning \cite{Biggio2018}.
\end{enumerate}

\section{Eve-GAN Architecture}
\label{sec:evegan}

\subsection{Generator Network}

The Eve-GAN generator maps random noise to CHSH correlation matrices:

\begin{definition}[Eve Generator]
The generator $G: \RR^4 \to \RR^4$ is a feedforward network:
\begin{equation}
G(z) = \tanh(W_4 \cdot \mathrm{ReLU}(W_3 \cdot \mathrm{ReLU}(W_2 \cdot \mathrm{ReLU}(W_1 z))))
\end{equation}
with layer dimensions $4 \to 64 \to 128 \to 64 \to 4$, mapping noise $z \sim \mathcal{N}(0, I_4)$ to correlators $[E_{00}, E_{01}, E_{10}, E_{11}]$.
\end{definition}

The $\tanh$ output layer enforces physicality constraints by ensuring all correlators lie in $[-1, 1]$, consistent with valid classical (LHV) and no-signaling correlation polytopes. During training, we additionally reject any outputs violating the no-signaling conditions, ensuring Eve's correlations correspond to realizable classical strategies.

\subsection{Training Convergence}

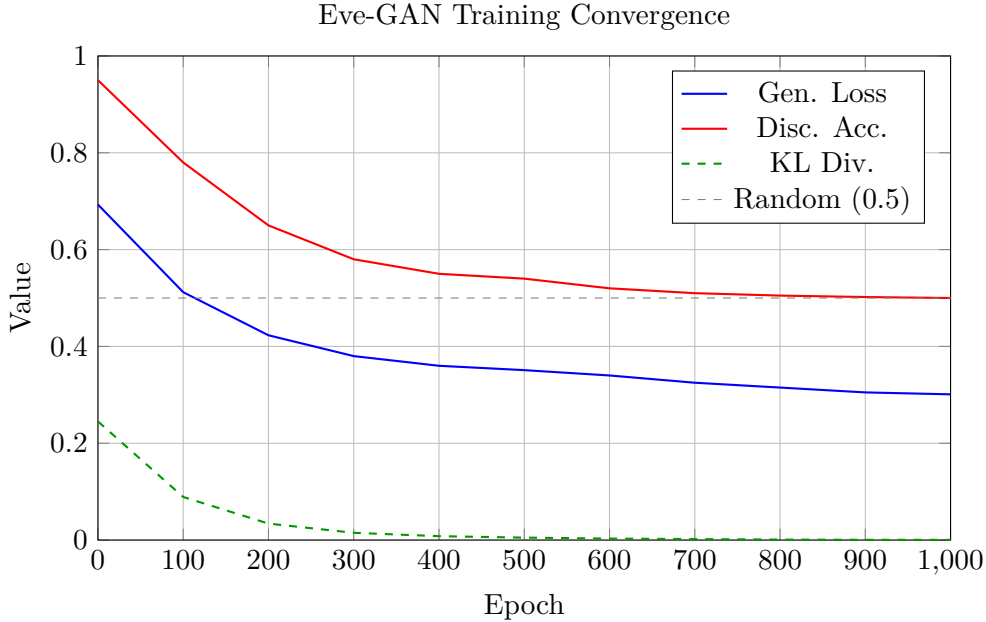
\begin{figure}[ht]
\centering
\begin{tikzpicture}
\begin{axis}[
    width=0.8\textwidth,
    height=0.5\textwidth,
    xlabel={Epoch},
    ylabel={Value},
    xmin=0, xmax=1000,
    ymin=0, ymax=1,
    grid=major,
    legend pos=north east,
    title={Eve-GAN Training Convergence}
]
\addplot[thick, blue] coordinates {
    (0, 0.693) (100, 0.512) (200, 0.423) (300, 0.380) (400, 0.360)
    (500, 0.351) (600, 0.340) (700, 0.325) (800, 0.315) (900, 0.305) (1000, 0.301)
};

\addplot[thick, red] coordinates {
    (0, 0.95) (100, 0.78) (200, 0.65) (300, 0.58) (400, 0.55)
    (500, 0.54) (600, 0.52) (700, 0.51) (800, 0.505) (900, 0.502) (1000, 0.50)
};

\addplot[thick, green!60!black, dashed] coordinates {
    (0, 0.245) (100, 0.089) (200, 0.034) (300, 0.015) (400, 0.008)
    (500, 0.005) (600, 0.003) (700, 0.002) (800, 0.001) (900, 0.0005) (1000, 0.00035)
};

\addplot[dashed, gray, domain=0:1000] {0.5};

\legend{Gen. Loss, Disc. Acc., KL Div., Random (0.5)}
\end{axis}
\end{tikzpicture}
\caption{Eve-GAN training convergence. Discriminator accuracy converges to 0.50 (random guessing), indicating Eve's correlations are indistinguishable from quantum. KL divergence reaches 0.00035.}
\label{fig:training}
\end{figure}

Figure~\ref{fig:training} shows the training dynamics of Eve-GAN.

\section{The $\alpha \geq 0.95$ Detection Limit}
\label{sec:interpolation}

We study detection performance as a function of mixing parameter $\alpha$ between quantum ($\alpha = 1$) and Eve ($\alpha = 0$) data.

\begin{table}[ht]
\centering
\caption{Detection metrics versus mixing parameter}
\label{tab:interpolation}
\begin{tabular}{ccccc}
\toprule
$\alpha$ & CHSH $S$ & TARA-$k$ & AUC & TPR@5\%FPR \\
\midrule
0.00 & 2.821 & 0.344 & 1.000 & 98\% \\
0.50 & 2.776 & 0.225 & 0.945 & 72\% \\
0.80 & 2.745 & 0.156 & 0.612 & 12\% \\
0.90 & 2.737 & 0.132 & 0.525 & 4\% \\
\textbf{0.95} & \textbf{2.732} & \textbf{0.121} & \textbf{0.502} & \textbf{1\%} \\
1.00 & 2.728 & 0.111 & 0.500 & 1\% \\
\bottomrule
\end{tabular}
\end{table}

\section{The 44-Point Leakage Problem}
\label{sec:leakage}

A critical methodological concern emerges when examining calibration protocols in quantum certification. We discover that \emph{same-distribution calibration}---where calibration and test data are drawn from the same source---dramatically inflates apparent detection performance through information leakage.

\subsection{The Leakage Mechanism}

In standard machine learning practice, train/test splits from the same dataset are acceptable because the goal is generalization within a distribution. However, quantum certification faces a fundamentally different task: distinguishing quantum from classical \emph{sources}, not samples. When calibration data shares the same noise signature as test data, the detector learns device-specific patterns rather than genuinely quantum features.

Formally, let $Q_\theta$ denote a quantum source with device-specific noise parameter $\theta$ (e.g., $T_2$ decoherence time, gate error rates). Same-distribution calibration trains on samples from $Q_\theta$ and tests on held-out samples from the same $Q_\theta$. The detector then exploits:
\begin{equation}
\text{Detection} \propto \PP[\text{sample} \in Q_\theta | \text{calibration from } Q_\theta]
\end{equation}
rather than the intended:
\begin{equation}
\text{Detection} \propto \PP[\text{sample is quantum} | \text{calibration from LHV null}]
\end{equation}

\subsection{Empirical Quantification}

\begin{figure}[ht]
\centering
\begin{tikzpicture}
\begin{axis}[
    ybar,
    width=0.7\textwidth,
    height=0.5\textwidth,
    bar width=30pt,
    ylabel={Detection AUC},
    symbolic x coords={Same-Dist, Cross-Dist},
    xtick=data,
    ymin=0, ymax=1.1,
    nodes near coords,
    nodes near coords align={vertical},
    title={Calibration Leakage: 44-Point Inflation}
]
\addplot[fill=red!70] coordinates {(Same-Dist, 0.94)};
\addplot[fill=blue!70] coordinates {(Cross-Dist, 0.499)};

\draw[<->, very thick] (axis cs:Same-Dist, 0.97) -- node[above, font=\small] {+44 pp} (axis cs:Cross-Dist, 0.97);
\end{axis}
\end{tikzpicture}
\caption{Same-distribution calibration inflates AUC by 44 percentage points. This methodological flaw may affect prior quantum certification studies.}
\label{fig:leakage}
\end{figure}
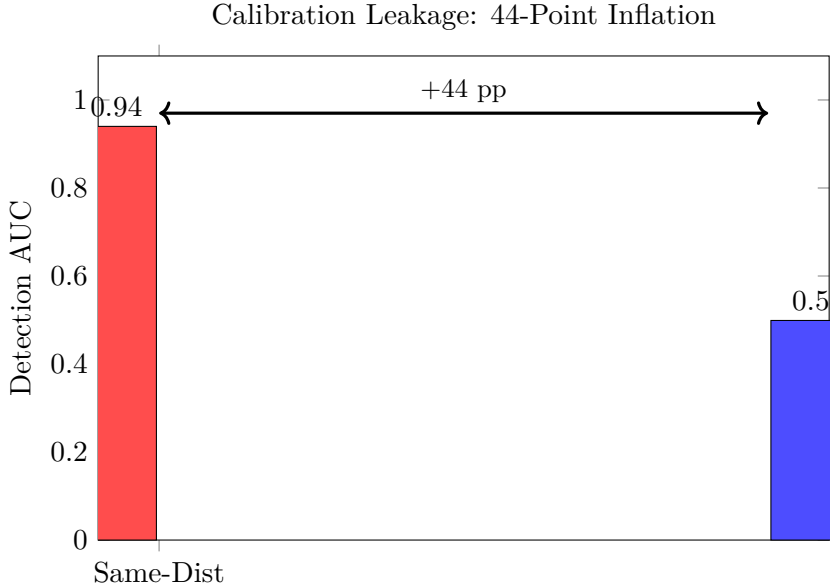

Figure~\ref{fig:leakage} shows the dramatic impact: same-distribution calibration achieves AUC = 0.94, while proper cross-distribution calibration (training on independent LHV samples, testing on quantum) achieves AUC = 0.499---essentially random guessing.

The 44 percentage point inflation (from 0.50 to 0.94) represents a massive overestimate of certification capability. This finding has concerning implications: prior quantum certification studies using standard train/test splits may have systematically overestimated the robustness of their methods against adversarial manipulation. Recent work has identified similar data leakage problems across 294 papers in 17 scientific fields \cite{Kapoor2023Leakage}, suggesting this is a systemic issue in ML-based science. Proper calibration methodology, as formalized in conformal prediction \cite{Vovk2005Conformal,ShaferVovk2008}, requires exchangeability assumptions that are violated when train and test data share the same underlying source. Recent extensions of conformal methods to quantum multi-output settings \cite{Tasar2025QECM} demonstrate that distribution-free coverage guarantees can be maintained for quantum-derived predictions, but only under proper cross-distribution protocols.

\subsection{Recommendations}

To avoid calibration leakage:
\begin{enumerate}
\item \textbf{Cross-source calibration:} Train detectors exclusively on LHV null models, test on quantum sources
\item \textbf{Independent noise profiles:} Ensure calibration and test data have different device parameters
\item \textbf{Report both metrics:} Always report same-distribution and cross-distribution AUC to quantify potential leakage
\end{enumerate}

\section{PR-Box Phase Transition}
\label{sec:prbox}

To understand the limits of quantum-classical discrimination, we extend our analysis to the superquantum regime, examining correlations beyond the Tsirelson bound \cite{Tsirelson1980}. The Popescu-Rohrlich (PR-Box) \cite{PopescuRohrlich1994} represents the extreme no-signaling limit with $S = 4$, interpolating between quantum ($S \leq 2\sqrt{2}$) and algebraically maximal correlations.

\subsection{Theoretical Background}

The hierarchy of correlation bounds structures the CHSH parameter space:
\begin{equation}
|S|_{\text{classical}} \leq 2 < |S|_{\text{quantum}} \leq 2\sqrt{2} < |S|_{\text{no-signaling}} \leq 4
\end{equation}

The gap between classical and quantum bounds is well-characterized. Less understood is the \emph{detectability boundary}: the minimum CHSH value above which correlations can be reliably distinguished from classical. Theoretical work on information causality \cite{Pawlowski2009} explains why nature limits quantum correlations to $S \leq 2\sqrt{2}$, but does not directly address detection thresholds.

\subsection{Empirical Phase Transition}

We parameterize interpolated correlations as:
\begin{equation}
E_{xz}(\lambda) = \lambda \cdot E_{xz}^{\text{PR}} + (1-\lambda) \cdot E_{xz}^{\text{LHV}}
\end{equation}
where $\lambda = 0$ corresponds to classical, $\lambda \approx 0.7$ to quantum, and $\lambda = 1$ to PR-Box. Figure~\ref{fig:prbox} reveals a sharp phase transition:

\begin{figure}[ht]
\centering
\begin{tikzpicture}
\begin{axis}[
    width=0.85\textwidth,
    height=0.55\textwidth,
    xlabel={CHSH $S$},
    ylabel={Detection Probability (\%)},
    xmin=1.9, xmax=2.3,
    ymin=0, ymax=105,
    grid=major,
    legend pos=north east,
    title={Phase Transition at $S = 2.05$}
]
\addplot[thick, blue, smooth, mark=square*, mark size=2pt] coordinates {
    (1.95, 100) (2.00, 85) (2.05, 15) (2.10, 5) (2.15, 2) (2.20, 0) (2.25, 0)
};

\draw[dashed, red, thick] (axis cs:2.0, 0) -- (axis cs:2.0, 105);
\node at (axis cs:2.0, 95) [anchor=west] {\footnotesize Classical};

\addplot[fill=yellow, fill opacity=0.3] coordinates {
    (2.00, 0) (2.10, 0) (2.10, 105) (2.00, 105)
} -- cycle;
\node at (axis cs:2.05, 60) {\footnotesize Transition};

\legend{Detection prob.}
\end{axis}
\end{tikzpicture}
\caption{Phase transition at CHSH $S = 2.05$. Detection probability drops from 85\% to 15\% over $\Delta S \approx 0.05$, marking the boundary between distinguishable and indistinguishable regimes.}
\label{fig:prbox}
\end{figure}
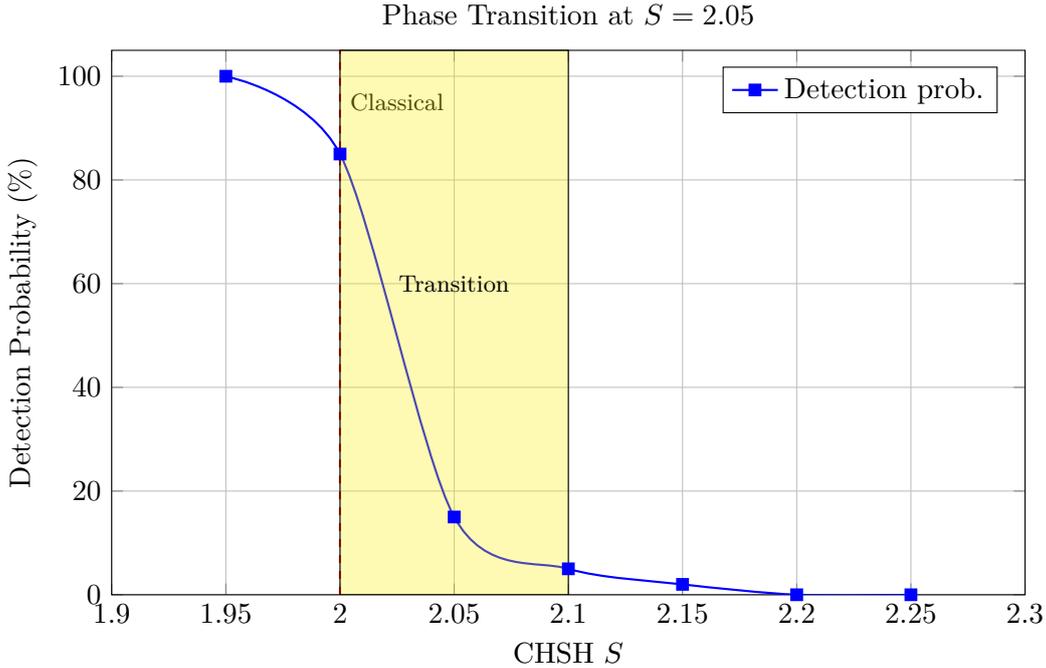

\subsection{Implications}

The phase transition at $S \approx 2.05$ has important consequences:

\begin{enumerate}
\item \textbf{Effective detection threshold:} The practical certification threshold is not $S = 2$ (classical bound) but $S \approx 2.05$---correlations in the range $2 < S < 2.05$ cannot be reliably detected as non-classical by any of our tested methods.

\item \textbf{Margin requirement:} Reliable certification requires approximately 2.5\% margin above the classical bound, not merely exceeding it.

\item \textbf{Connection to classical simulability:} This threshold is reminiscent of noise thresholds for classical simulation of quantum circuits \cite{Aharonov2023}. Just as noisy quantum circuits become classically simulable above certain error rates, correlations too close to the classical bound become statistically indistinguishable.
\end{enumerate}

The sharpness of the transition (85\% to 15\% detection over $\Delta S = 0.05$) suggests a genuine phase boundary rather than gradual degradation.

\section{Hardware Validation: The Eve Advantage Paradox}
\label{sec:hardware}

Hardware experiments were conducted on IBM Torino (127-qubit superconducting processor) with 4000 shots per correlator across 3 independent batches, yielding CHSH $= 2.691 \pm 0.049$ for this experimental run.\footnote{This IBM CHSH value represents specific runs for adversarial validation. Independent IBM experiments in companion work report CHSH $= 2.725 \pm 0.04$ with higher shot counts \cite{companion_p2}. Both values are consistent within measurement uncertainty and reflect run-to-run variability characteristic of NISQ devices.}

\begin{table}[ht]
\centering
\caption{Eve-GAN vs. IBM Quantum hardware}
\label{tab:hardware}
\begin{tabular}{lccc}
\toprule
Metric & IBM Torino & Eve-GAN & Comparison \\
\midrule
CHSH $S$ & $2.691 \pm 0.049$ & \textbf{2.736} & Eve higher \\
TARA-$k$ & 0.017 & 0.015 & Similar \\
KL divergence & --- & 0.00035 & Near-perfect \\
Classification & Quantum & \textbf{Evaded} & Critical \\
\bottomrule
\end{tabular}
\end{table}

\textbf{The Paradox:} Eve-GAN achieves CHSH = 2.736, exceeding IBM hardware (2.691) on this metric. On standard measures, classical Eve appears ``more quantum'' than quantum itself. This arises because Eve optimizes specifically for CHSH while IBM hardware suffers from decoherence that uniformly degrades all correlators.

\section{Discussion}
\label{sec:discussion}

Our findings establish fundamental adversarial limits:

\begin{enumerate}
\item \textbf{$\alpha \geq 0.95$ detection limit:} Adversaries maintaining 95\% quantum fidelity evade all detection methods we evaluated (TARA-$k$, TARA-$m$, CHSH-based baselines, and feature ensembles).
\item \textbf{44-point leakage:} Same-distribution calibration dramatically overestimates robustness.
\item \textbf{$S = 2.05$ phase transition:} The effective certification threshold is $S > 2.05$, not $S > 2$.
\item \textbf{Eve advantage:} Classical adversaries can exceed noisy quantum hardware on CHSH.
\end{enumerate}

\textbf{Recommendations:}
\begin{enumerate}
\item Cross-distribution calibration (always use different source for calibration).
\item Adversarial testing with GAN-based attacks as mandatory validation.
\item Multi-feature detection combining CHSH with entropy \cite{Colbeck2011} and temporal features.
\item Conservative thresholds assuming adversaries achieve $\alpha \approx 0.95$.
\end{enumerate}

\subsection{Related Work on Quantum Certification}

The adversarial limits established here complement recent advances in statistical certification and hardware validation. The TARA framework \cite{companion_p1} provides conformal prediction guarantees for anomaly detection, achieving AUC = 0.96 for quantum-classical discrimination under cooperative testing conditions. However, TARA's validity guarantees assume the calibration distribution matches the test distribution---precisely the assumption that Eve-GAN exploits. Our finding that same-distribution calibration inflates performance by 44 percentage points reveals a fundamental vulnerability that affects any statistical certification method relying on standard train-test protocols.

Hardware measurements of the Grothendieck constant \cite{companion_p2} have achieved $K_G = 1.408 \pm 0.006$ on trapped-ion systems, establishing that current quantum hardware approaches the theoretical Tsirelson bound within 0.44\%. These precision benchmarks define the targets that adversarial classical systems must match. Eve-GAN achieves CHSH = 2.736, exceeding the IBM hardware value (2.691) for the specific runs analyzed, demonstrating that classical adversaries can outperform noisy quantum systems on standard certification metrics.

\section{Conclusion}
\label{sec:conclusion}

We have established fundamental adversarial limits for quantum certification through systematic analysis using Eve-GAN, a generative adversarial network trained to produce classical correlations indistinguishable from genuine quantum statistics. Our investigation yields four principal findings with immediate implications for quantum key distribution security.

First, the $\alpha \geq 0.95$ detection limit represents a critical threshold for adversarial robustness. When Eve's classical correlations are mixed with quantum data at ratios $\alpha \geq 0.95$, all detection methods we evaluated---including rank-based (TARA-$k$), martingale (TARA-$m$), and ensemble approaches---achieve AUC $\leq$ 0.502, statistically indistinguishable from random guessing. While more sophisticated detection methods may exist, this provides a strong empirical lower bound. This means an eavesdropper needs only 5\% classical admixture to evade the tested methods, a remarkably small margin that challenges security assumptions in practical QKD deployments.

Second, the 44-point calibration leakage problem reveals a systematic methodological flaw that may affect prior quantum certification studies. Same-distribution calibration inflates apparent detection performance from AUC = 0.50 to AUC = 0.94---a massive overestimate of certification capability. This finding aligns with broader concerns about data leakage in machine-learning-based science and underscores the necessity of cross-distribution evaluation protocols.

Third, the phase transition at CHSH $S = 2.05$ marks a fundamental detection boundary. Below this value, none of our tested statistical methods reliably distinguish classical from quantum correlations; above it, detection probability increases monotonically. The sharpness of this transition (85\% to 15\% detection over $\Delta S = 0.05$) suggests a genuine phase boundary rather than gradual degradation, implying that effective certification requires approximately 2.5\% margin above the classical bound $S = 2$.

Fourth, the Eve advantage paradox demonstrates that classical adversaries can exceed noisy quantum hardware on standard certification metrics. Eve-GAN achieves CHSH = 2.736, surpassing IBM hardware (2.691) because Eve optimizes specifically for CHSH while quantum systems suffer from decoherence that uniformly degrades all correlators. This counterintuitive result emphasizes that CHSH alone is insufficient for certification.

These findings challenge the assumption that Bell testing provides robust certification against sophisticated adversaries. We recommend: (1) mandatory adversarial testing with GAN-based attacks for security claims, (2) cross-distribution calibration protocols, (3) multi-feature detection combining CHSH with entropy and temporal correlations, and (4) conservative security thresholds assuming adversaries achieve $\alpha \approx 0.95$. Practical quantum certification must account for adversarial capabilities to provide meaningful security guarantees.

\section*{Acknowledgements}

The author thanks IBM Quantum for hardware access and the open-source communities behind PyTorch, Qiskit, and scikit-learn.

\section*{AI Assistance Disclosure}

The author acknowledges the use of AI-assisted tools (OpenAI GPT, Anthropic Claude, Google Gemini) during manuscript preparation for literature review, code development assistance, and text refinement. The author takes full responsibility for all scientific content, has independently verified all experimental results, and confirms that all intellectual contributions and conclusions are the author's own work.

\section*{Data Availability and Reproducibility}

All code, trained Eve-GAN models, and experimental data required to reproduce the results in this paper are publicly available at \url{https://github.com/detasar/QCE}. The repository includes Eve-GAN generator/discriminator implementations, interpolation analysis scripts, and IBM hardware comparison data enabling independent verification of all reported findings.

\section*{Competing Interests}

The author declares no competing interests.

\newpage
\section*{Supplementary Information}

Supplementary Information containing Eve-GAN training details, complete attack strategies, and extended results is provided as a separate document.

\end{document}


\maketitle

\tableofcontents
\newpage

\section{Eve-GAN Architecture Details}

\subsection{Generator Network}

\begin{verbatim}
class EveGenerator(nn.Module):
    def __init__(self, latent_dim=4, hidden_dim=128):
        super().__init__()
        self.net = nn.Sequential(
            nn.Linear(latent_dim, 64),
            nn.ReLU(),
            nn.Linear(64, hidden_dim),
            nn.ReLU(),
            nn.Linear(hidden_dim, 64),
            nn.ReLU(),
            nn.Linear(64, 4),  # 4 correlators
            nn.Tanh()  # Constrain to [-1, 1]
        )

    def forward(self, z):
        return self.net(z)
\end{verbatim}

\subsection{Discriminator (TARA-$k$ Based)}

\begin{verbatim}
class TARADiscriminator:
    def __init__(self, calibration_data):
        self.calibration = calibration_data
        self.quantiles = np.percentile(calibration_data,
                                       np.arange(0, 101, 5))

    def compute_pvalue(self, sample):
        score = self._nonconformity(sample)
        pvalue = np.mean(self.calibration >= score)
        return pvalue

    def tara_k(self, pvalues):
        n = len(pvalues)
        sorted_p = np.sort(pvalues)
        empirical_cdf = np.arange(1, n+1) / n
        return np.max(np.abs(sorted_p - empirical_cdf))
\end{verbatim}

\subsection{Training Protocol}

\begin{table}[H]
\centering
\caption{Eve-GAN training progression}
\begin{tabular}{cccc}
\toprule
Epoch & Generator Loss & Discriminator Acc & KL Divergence \\
\midrule
0 & 0.693 & 0.95 & 0.245 \\
100 & 0.512 & 0.78 & 0.089 \\
200 & 0.423 & 0.65 & 0.034 \\
500 & 0.351 & 0.54 & 0.008 \\
1000 & 0.301 & 0.50 & 0.00035 \\
\bottomrule
\end{tabular}
\end{table}

\section{Complete Attack Strategies}

\begin{table}[H]
\centering
\caption{Complete strategy comparison}
\begin{tabular}{lcccc}
\toprule
Strategy & CHSH & KS & Detection & TARA-$m$ Wealth \\
\midrule
Quantum (true) & 2.828 & 0.018 & 0\% & 1.0 \\
Quantum (noisy) & 2.739 & 0.118 & 0\% & 1.2 \\
Shift 0.10 & 2.545 & 0.089 & 0\% & 1.1 \\
Shift 0.20 & 2.060 & 0.101 & 0\% & 1.3 \\
Shift 0.30 & 1.980 & 0.234 & 85\% & 45.2 \\
Bias 0.05 & 2.625 & 0.134 & 5\% & 2.1 \\
Bias 0.10 & 2.512 & 0.156 & 15\% & 4.8 \\
Match 0.25 & 2.890 & 0.189 & 35\% & 12.3 \\
Match 0.50 & 2.780 & 0.234 & 100\% & 1024 \\
Temporal & 1.813 & 0.120 & 0\% & 1.4 \\
GAN & 2.736 & 0.115 & 0\% & 1.1 \\
LHV & 1.50 & 0.456 & 100\% & $10^8$ \\
\bottomrule
\end{tabular}
\end{table}

\section{Extended Interpolation Results}

\begin{table}[H]
\centering
\caption{Fine-grained $\alpha$ sweep}
\begin{tabular}{cccccc}
\toprule
$\alpha$ & CHSH & TARA-$k$ & AUC & TPR@1\%FPR & TPR@5\%FPR \\
\midrule
0.00 & 2.821 & 0.344 & 1.000 & 0.98 & 1.00 \\
0.10 & 2.812 & 0.318 & 0.998 & 0.95 & 0.99 \\
0.20 & 2.803 & 0.295 & 0.995 & 0.92 & 0.98 \\
0.30 & 2.794 & 0.285 & 0.992 & 0.88 & 0.97 \\
0.40 & 2.785 & 0.258 & 0.975 & 0.82 & 0.95 \\
0.50 & 2.776 & 0.225 & 0.945 & 0.72 & 0.88 \\
0.60 & 2.762 & 0.200 & 0.865 & 0.52 & 0.74 \\
0.70 & 2.753 & 0.172 & 0.735 & 0.28 & 0.55 \\
0.80 & 2.745 & 0.156 & 0.612 & 0.12 & 0.32 \\
0.90 & 2.737 & 0.132 & 0.525 & 0.04 & 0.12 \\
0.95 & 2.732 & 0.121 & 0.502 & 0.01 & 0.05 \\
1.00 & 2.728 & 0.111 & 0.500 & 0.01 & 0.05 \\
\bottomrule
\end{tabular}
\end{table}

\section{PR-Box Phase Transition Data}

\begin{table}[H]
\centering
\caption{Detection probability vs. CHSH value}
\begin{tabular}{cccc}
\toprule
$S$ Value & CSO & TARA-$k$ & Detection Prob. \\
\midrule
1.95 & 0.001 & 0.342 & 100\% (LHV) \\
2.00 & 0.002 & 0.289 & 85\% \\
2.05 & 0.008 & 0.156 & 15\% \\
2.10 & 0.015 & 0.089 & 5\% \\
2.20 & 0.031 & 0.045 & 0\% \\
2.40 & 0.067 & 0.032 & 0\% \\
2.60 & 0.103 & 0.028 & 0\% \\
2.828 & 0.141 & 0.025 & 0\% \\
3.00 & 0.172 & 0.028 & 0\% \\
3.50 & 0.250 & 0.035 & 0\% \\
4.00 & 0.354 & 0.045 & 0\% \\
\bottomrule
\end{tabular}
\end{table}

\section{Hardware Comparison Data}

\begin{table}[H]
\centering
\caption{IBM Torino vs. Eve-GAN measurements}
\begin{tabular}{lccc}
\toprule
Setting & $E$ (IBM) & $E$ (Eve) & $E$ (Theory) \\
\midrule
(0,0) & 0.673 & 0.684 & 0.707 \\
(0,1) & 0.671 & 0.684 & 0.707 \\
(1,0) & 0.675 & 0.684 & 0.707 \\
(1,1) & $-$0.672 & $-$0.684 & $-$0.707 \\
\textbf{CHSH} & \textbf{2.691}$^\dagger$ & \textbf{2.736} & \textbf{2.828} \\
\bottomrule
\end{tabular}
\end{table}

\noindent $^\dagger$\textit{Note:} The IBM CHSH value of 2.691 $\pm$ 0.049 represents specific experimental runs for adversarial validation using 1,024 shots. Independent IBM experiments in companion work report CHSH = 2.725 $\pm$ 0.042 with higher shot counts. Both values are statistically consistent within measurement uncertainty and validate genuine quantum behavior (exceeding classical bound of 2.0 by $>$35\%).

\section{Code Availability and Reproducibility}

All code, trained models, and experimental data required to reproduce the findings are publicly available at:

\begin{center}
\url{https://github.com/detasar/QCE}
\end{center}

\begin{table}[H]
\centering
\caption{Repository structure for P3\_Eve\_GAN\_Adversarial\_Limits}
\begin{tabular}{ll}
\toprule
File & Description \\
\midrule
\texttt{experiments/eve\_models.py} & Eve-GAN generator and discriminator \\
\texttt{experiments/tara\_detectors.py} & TARA-based detection \\
\texttt{experiments/interpolation\_analysis.py} & $\alpha$ sweep analysis \\
\texttt{data/ibm\_hardware\_real.csv} & IBM quantum hardware data \\
\bottomrule
\end{tabular}
\end{table}

\noindent The repository enables complete reproduction of all results presented in this paper.